\documentclass[printer]{aa}

\usepackage{graphicx}
\usepackage{txfonts}
\usepackage[colorlinks,citecolor=blue]{hyperref}
\begin{document}

   \title{Infrared diagnostics of the solar magnetic field with Mg I 12 $\mu$m lines: forward-model results}


   \author{Xin Li\inst{1,2}
          \and YongLiang Song\inst{1}
          \and H. Uitenbroek\inst{3}
          \and Xiao Yang\inst{1}
          \and XianYong Bai\inst{1,2}
          \and YuanYong Deng\inst{1,2}}


   \institute{Key Laboratory of Solar Activity, National Astronomical Observatories, Chinese Academy of Sciences, 20A Datun Road, Beijing 100101, People's Republic of China
             \email{xybai@bao.ac.cn}
         \and
             School of Astronomy and Space Science, University of Chinese Academy of Sciences, No.19(A) Yuquan Road, Beijing 100049, People's Republic of China
         \and
             National Solar Observatory, University of Colorado Boulder, 3665 Discovery Drive, Boulder, CO 80303, USA}


  \abstract
   {The Mg I 12.32 and 12.22 $\mu$m lines are a pair of emission lines that present a great advantage for accurate solar magnetic field measurement. They potentially contribute to the diagnosis of solar atmospheric parameters through their high magnetic sensitivity.}
   {The goal of this study is to understand the radiation transfer process of these lines in detail and explore the ability of magnetic field diagnosis in the infrared.}
   {We calculated the Stokes profiles and response functions of the two Mg I 12 $\mu$m lines based on one-dimensional solar atmospheric models using the Rybicki-Hummer (RH) radiative transfer code. The integration of these profiles with respect to the wavelength was used to generate calibration curves related to the longitudinal and transverse fields. The traditional single-wavelength calibration curve based on the weak-field approximation was also tested to determine if it is suitable for the infrared.}
   {The 12.32 $\mu$m line is more suitable for a magnetic field diagnosis because its relative emission intensity and polarization signal are stronger than that of the 12.22 $\mu$m line. The result from the response functions illustrates that the derived magnetic field and velocity with 12.32 $\mu$m line mainly originate from the height of 450 km, while that for the temperature is about 490 km. The calibration curves obtained by the wavelength-integrated method show a nonlinear distribution. For the Mg I 12.32 $\mu$m line, the longitudinal (transverse) field can be effectively inferred from Stokes V/I (Q/I and U/I) in the linear range below $\sim 600$ G ($\sim 3000$ G) in quiet regions and below $\sim 400$ G ($\sim 1200$ G) in penumbrae. Within the given linear range, the method is a supplement to the magnetic field calibration when the Zeeman components are incompletely split.}
   {}


   \titlerunning{Solar magnetic field diagnostics with Mg I 12 $\mu$m lines: forward-model results}
   \authorrunning{Xin Li et al.}

   \maketitle
%

\section{Introduction}

Magnetic fields play a significant role in various solar activities. To obtain more accurate magnetic field values, infrared Stokes polarimetry has been attempted in recent decades, given the high magnetic sensitivity of lines in the infrared solar spectrum \citep{2014Matthew}. The Cryogenic Infrared Spectrograph instrument \citep{2010Cao} of the 1.6 m Goode Solar Telescope at Big Bear Solar Observatory and the Cryogenic Near-Infrared Spectro-Polarimeter instruments \citep{2016Woeger} of the 4 m Daniel K. Inouye Solar Telescope at the National Solar Observatory have been designed to measure the solar magnetic field in the 1--5 $\mu$m infrared band. Of the many magnetic sensitive lines in the infrared, the Mg I emission lines at 12.32 and 12.22 $\mu$m (or 811.575 and 818.058 cm$^{-1}$, hereafter Mg I 12 $\mu$m lines) were found to have the highest magnetic sensitivity so far. They were first reported by \cite{1981Murcray} and then identified as transitions between high Rydberg levels of the Mg I atom by \cite{1983Chang&Noyes}. These lines show clear Zeeman splitting when the magnetic field strength is only a few hundred Gauss \citep{1983Brault&Noyes}. They have a great potential for magnetic field measurements, especially for the detection of the weak magnetic field.

There are many important studies in the early observation and modeling of the Mg I 12 $\mu$m lines. In terms of observation, \cite{1983Brault&Noyes} found that these lines with extremely narrow emission peaks and wide absorption troughs show limb brightening, and they usually appear in quiet regions, sunspot penumbrae, and plages, but are absent in sunspot umbrae. The high-resolution spectrum observation of the Mg I 12.32 $\mu$m line shows velocity oscillations, but no intensity oscillations \citep{1988Deming,1991Deming}. For modeling, \cite{1992Carlsson} successfully synthesized the line profiles with an excellent agreement with the observations for the first time by considering nonlocal thermodynamic equilibrium (NLTE). They confirmed that these lines were formed in the photosphere instead of the chromosphere and that population departure divergence of the high Rydberg levels is the reason for the formation of the emergent line profiles. This was an important contribution to understanding the formation mechanism of the Mg I 12 $\mu$m lines. The observed peak-and-trough limb brightening has also been simulated and reproduced. Since then, people can carry out quantitative modeling and diagnostic application for these lines. \cite{1995Bruls} synthesized the polarization profiles of the Mg I 12 $\mu$m lines and found that the formation heights of the emission features in quiet regions and sunspot penumbrae are approximately the same through the contribution function. Recently, \cite{2020Hong} performed NLTE calculations on the Mg I 12.32 $\mu$m line under flare atmosphere models, and the results showed that the change in the line formation height from the upper photosphere to the chromosphere during flare heating can cause the Zeeman splitting width and the Stokes V lobe intensity to decrease.

The two Mg I 12 $\mu$m lines have a very high magnetic sensitivity, but it is still unclear what their potential contribution in the diagnosis of solar atmospheric parameters might be, for instance, magnetic field and temperature. There are two main approaches to deriving magnetic fields. One is calculating the magnetic fields directly from the Zeeman splitting, but this cannot be used when the Zeeman components are incompletely separated \citep{2001Jennings}. The other method is inverting the magnetic fields from the Stokes profiles based on the radiative transfer theory \citep{1993Hewagama}. The inversion based on the local thermodynamic equilibrium (LTE) is represented by the Milne-Eddington inversion method, which proved to be quite robust and relatively stable \citep{1987Skumanich}. However, the NLTE-based inversion is a time-consuming process. Furthermore, these inversion procedures with multiple free parameters usually face problems of convergence and uniqueness of the solutions.

In this paper, we perform NLTE calculations of the two Mg I 12 $\mu$m lines using quiet-Sun and sunspot penumbra atmospheric models and compare their differences in the magnetic field diagnosis. Based on the result, we also try to find a new calibration method that can infer the magnetic field in a short time without using inversion when the magnetic field is so weak that the Zeeman components are unresolved. The feasibility of measuring the magnetic field by employing a filter-based magnetograph observed at the single-wavelength point of the Mg I 12 $\mu$m lines is also discussed. This work is therefore helpful not only for the future of solar magnetic field telescopes working at mid-infrared wavelengths, but also for the Infrared System for the Accurate Measurement of Solar Magnetic Field (AIMS) under construction in China, which selected the Mg I 12 $\mu$m lines as its main working lines \citep{2016Deng}.

The paper is organized as follows. In Section 2 we introduce the methods and models, including the numerical approach and the atomic and atmospheric models, followed by the results in Section 3. Finally, in Section 4 we summarize and discuss our main results, including an outlook on future research.
\section{Methods and models}
For the Mg I 12 $\mu$m emission lines, we must take optically thick radiative transfer into account. To understand the radiation transfer process of these lines, a comprehensive atomic model must be selected to perform detailed numerical radiative transfer calculations under a given atmospheric model. The Rybicki-Hummer (RH) code is the numerical radiative transfer code based on the Multi-level Approximate Lambda Iteration formalism of \cite{1991Rybicki,1992Rybicki}, which was later improved by Uitenbroek \citep{2001Uitenbroek,2015Pereira&Uitenbroek}. Considering the advantages of the RH code in calculating NLTE radiation transfer, we used it to calculate the detailed radiative transfer of the Mg I 12 $\mu$m lines. In radiative transfer calculations, radiative transfer equations determine the absorption and emission state of the radiation in the solar atmosphere, and statistical equilibrium equations give the populations at each energy level of the atom, which are coupled with each other and can be solved self-consistently. We assumed complete frequency redistribution, and the Zeeman effect and the magneto-optic (M-O) effect were considered as well. As a result, we obtained polarization spectra of the Mg I 12 $\mu$m lines in different solar atmosphere models with different magnetic field strengths at disk center by setting $\mu=1.0$ (viewing angles are indicated by $\mu=\cos\theta$, and $\mu=1.0$ means the disk center). The input magnetic field strength is a fixed value and does not change with height.

We used the model atom \textsf{Mg I\_66.atom}, which is based on the atomic model of \cite{1992Carlsson}, which  properly considers all population processes. It is well known that the Mg atom has many energy levels, and the transitions between the energy levels are very complicated. The 12.32 and 12.22 $\mu$m lines are identified as $3s7i^{1,3}I^{e}-3s6h^{1,3}H^{0}$ and $3s7h^{1,3}H^{0}-3s6g^{1,3}G^{e}$ transitions of Mg I, respectively \citep{1983Chang&Noyes}. \cite{1987Chang} deduced that their Land\'e g-factor was unity, which was subsequently verified by \cite{1988Lemoine} in a laboratory measurement. Because the Mg I 12 $\mu$m lines are formed in NLTE \citep{1992Carlsson}, we need to consider enough energy levels to better model the emission features consistent with the observations.

\begin{figure*}[!htbp]
\centering
\includegraphics[width=18cm, angle=0]{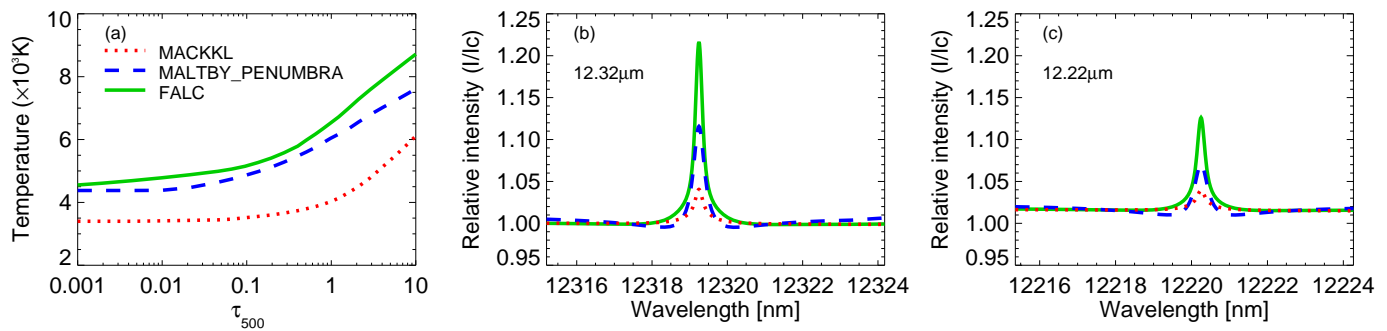} \caption{Different solar atmosphere models and corresponding synthetic Stokes I profiles. Panel (a) Temperature against the continuum optical depth at $\lambda$ = 500 nm for the three solar atmosphere models. Panels (b) and (c) are synthetic Stokes I profiles of the Mg I 12.32 $\mu$m and Mg I 12.22 $\mu$m without magnetic fields. The relative intensity profiles are relative to the continuum intensity in the corresponding model.}
\label{fig1}
\end{figure*}

We chose the sunspot penumbra, umbra, and also quiet-Sun atmosphere models to study these lines. Figure~\ref{fig1}.a shows the temperature stratifications of the three different solar atmosphere models. They are all standard one-dimensional plane-parallel atmosphere models, which means that all the physical parameters only change with depth in the atmosphere. For different atmospheric models, the formation of the Mg I 12 $\mu$m lines is different. The FALC in Figure~\ref{fig1}.a is a static model of the quiet Sun proposed by \cite{1993Fontenla}. The second model, labeled MACKKL \citep{1986Maltby}, is a sunspot umbra atmosphere model, and the third model named MALTBY\_PENUMBRA \citep[hereafter MALTBY]{1969Kjeldseth} is a sunspot penumbra model. The largest difference between the MALTBY model and other two models is that it does not possess a chromospheric temperature rise.

\section{Results}
\subsection{Synthetic Stokes profiles}
Figure~\ref{fig1}.b and Figure~\ref{fig1}.c show the synthetic Stokes I profiles of the Mg I 12.32 $\mu$m and Mg I 12.22 $\mu$m lines without a magnetic field at disk center for the three solar atmosphere models, respectively. The results show that the synthetic profiles of two Mg I 12 $\mu$m lines agree with the observed profiles \citep{1992Carlsson}. The observed emission peaks, the wide absorption troughs, and the observed strength ratios are reproduced quiet well. The synthetic profiles in the MACKKL model also show weak emissions, but early observations by \cite{1983Brault&Noyes} showed that the 12 $\mu$m emission lines are normally absent in the umbrae. The reason might be the Saha-Boltzmann temperature sensitivity \citep{1995Bruls}. Normally, the 12 $\mu$m emission lines are hard to observe because the low temperature of sunspot umbra leads to a low population density in the high Rydberg states, resulting in weak radiation. However, there may be emission if the temperature is high enough to cause a slight difference between the population departures of the upper and lower levels of the Mg I 12 $\mu$m lines. For example, \cite{1990Deming} found that flare heating can excite emission lines in the umbrae. Nonetheless, there may be no corresponding observations to verify the simulation results obtained in the umbra atmosphere. The FALC and MALTBY models are therefore included in the following calculations, and the MACKKL model is not considered. The relative intensity of the 12.32 $\mu$m line in the three atmospheric models is also stronger than that of 12.22 $\mu$m line.

\begin{figure*}[!htbp]
\centering
\includegraphics[width=18cm, angle=0]{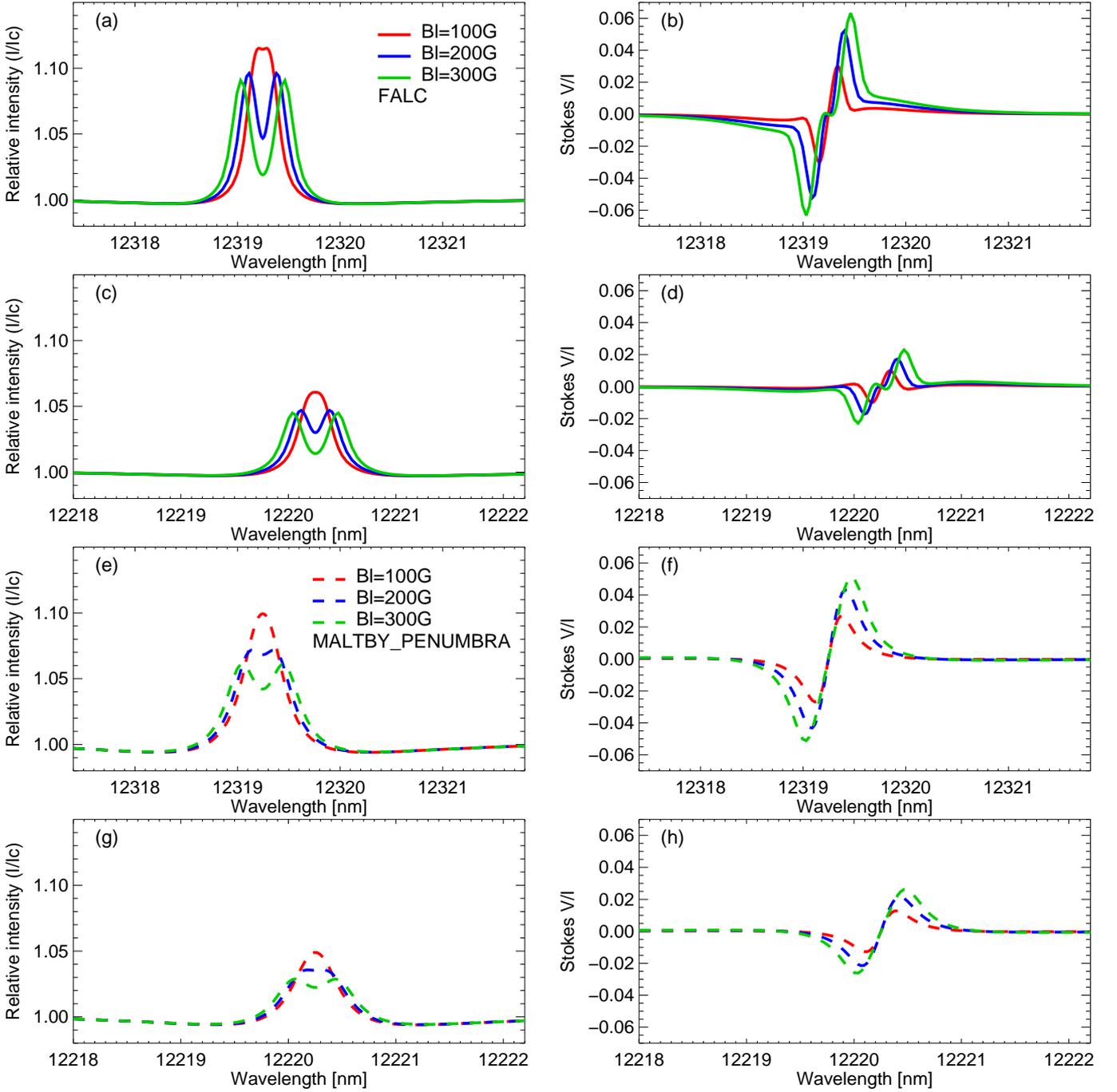} \caption{Calculated Stokes I (left panels) and V (right panels) profiles of two Mg I 12 $\mu$m lines for the quiet Sun (solid, top four panels) and sunspot penumbrae (dashed, bottom four panels) models with longitudinal magnetic fields (i.e., inclination angle equals to $ 0^{\circ}$, expressed by $B_l$). Panels (a), (b), (e), and (f) and panels (c), (d), (g), and (h) correspond to the Mg I 12.32 $\mu$m and Mg I 12.22 $\mu$m lines, respectively. Different colours are for different magnetic field strength. Red, blue, and green lines correspond to magnetic field strengths of $B_l$ = 100 G, $B_l$ = 200 G, and $B_l$ = 300 G, respectively. The relative intensity profiles are relative to the continuum intensity.}
\label{fig2}
\end{figure*}

\begin{figure*}[!htbp]
\centering
\includegraphics[width=18cm, angle=0]{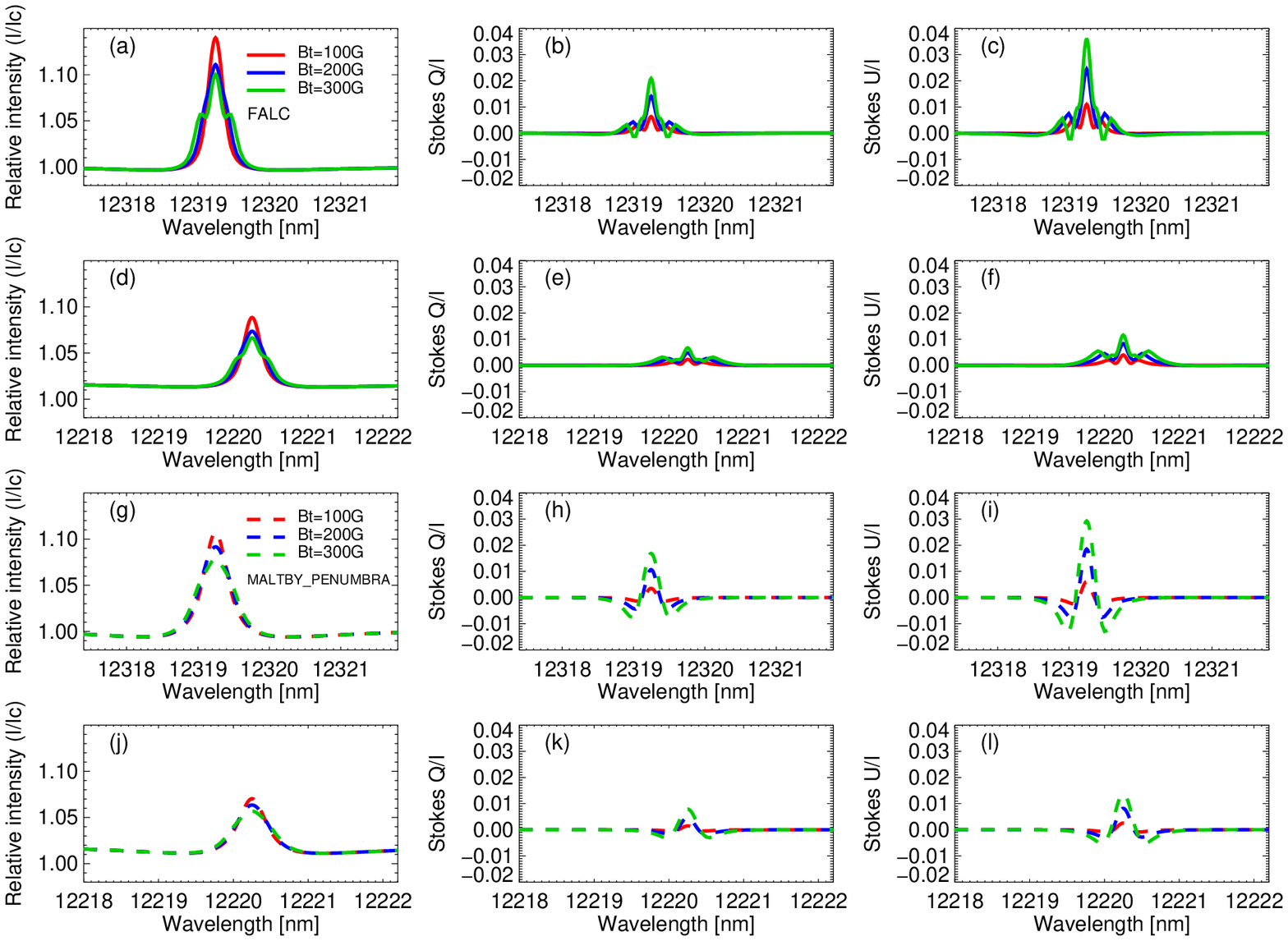} \caption{Same as Fig. 2, but for the transverse magnetic fields (expressed by $B_t$), and the azimuth angle equals $\chi= 30^{\circ}$. The left, middle, and right panels show Stokes I, Q, and U profiles, respectively. Panels (a), (b), (c), (g), (h), and (i) and panels (d), (e), (f), (j), (k), and (l) correspond to the Mg I 12.32 $\mu$m and Mg I 12.22 $\mu$m lines, respectively. The relative intensity profiles are relative to the continuum intensity.}
\label{fig3}
\end{figure*}

Next, we calculated the splitting behavior of the two Mg I 12 $\mu$m lines by adding magnetic fields. Figure~\ref{fig2} (Figure~\ref{fig3}) displays the Stokes profiles for the FALC and MALTBY models with $B_l$ ($B_t$) of 100, 200, and 300 G. The Stokes I, Q, U, and V profiles of the two 12 $\mu$m lines are similar. However, the average polarization signal of the 12.32 $\mu$m line with different magnetic field strengths is three times stronger than that of the 12.22 $\mu$m line in the FALC model, and twice as strong in the MALTBY model. The feature in the center of the Stokes V profile in the quiet Sun with 300 G appears to be an M-O reversal (Figure~\ref{fig2} d). For comparison, we calculated the Stokes V profile without considering the M-O effect for the same conditions, and the results show that this feature still exists. This means that the feature in the center of the Stokes V profile is not caused by the M-O effect. The Stokes Q and U profiles primarily show the emission parts and are more complicated (with many wiggles, or inflections) in the quiet regions than in the penumbrae (Figure~\ref{fig3}). This result can be explained by the more obvious absorption troughs and the wider line width of the Stokes I profiles in the penumbra model.


Furthermore, the Mg I 12 $\mu$m lines show a different splitting behavior for different magnetic field strengths. The first columns of Figures~\ref{fig2} and~\ref{fig3} show that as the magnetic field strength increases, the Stokes I profiles exhibit a very clear Zeeman splitting, and the line width also gradually increases. In the FALC model, the 12.32 $\mu$m line splits into two $\sigma$ components at $B_l$ = 100 G, while the 12.22 $\mu$m line only splits at $B_l$ = 200 G (Figure~\ref{fig2} a,c). Correspondingly, in the MALTBY model, the 12.32 $\mu$m line shows a split at $B_l$ = 200 G, while the 12.22 $\mu$m line only shows a split at $B_l$ = 300 G (Figure~\ref{fig2} e,g). The two Mg I 12 $\mu$m lines split into two $\sigma$ components and a $\pi$ component at $B_t$ = 300 G in the FALC model (Figure~\ref{fig3} a,d), but the lines do not show a split even at $B_t$ = 300 G in the MALTBY model (Figure~\ref{fig3} g,j). We conclude that a higher magnetic field strength is required in the penumbrae than in quiet regions to see the split due to the greater line width in penumbrae. The larger line width in the penumbrae may be caused by increased pressure broadening, because for the cool model, the Mg I 12 $\mu$m lines are formed deeper and the density there is higher \citep{1995Bruls}. The 12.32 $\mu$m line splits at a lower magnetic field than the 12.22 $\mu$m line. The intensity of the line center and continuum of the 12.32 $\mu$m line is also stronger than that of the 12.22 $\mu$m line. Therefore, although their Land\'e g-factors are the same, the polarization signals of Q/I, U/I, and V/I of 12.32 $\mu$m are greater than those of the 12.22 $\mu$m line.
\subsection{Response function from Stokes I to the magnetic field, temperature, and velocity}
\begin{figure*}[!htbp]
\centering
\includegraphics[width=18cm, angle=0]{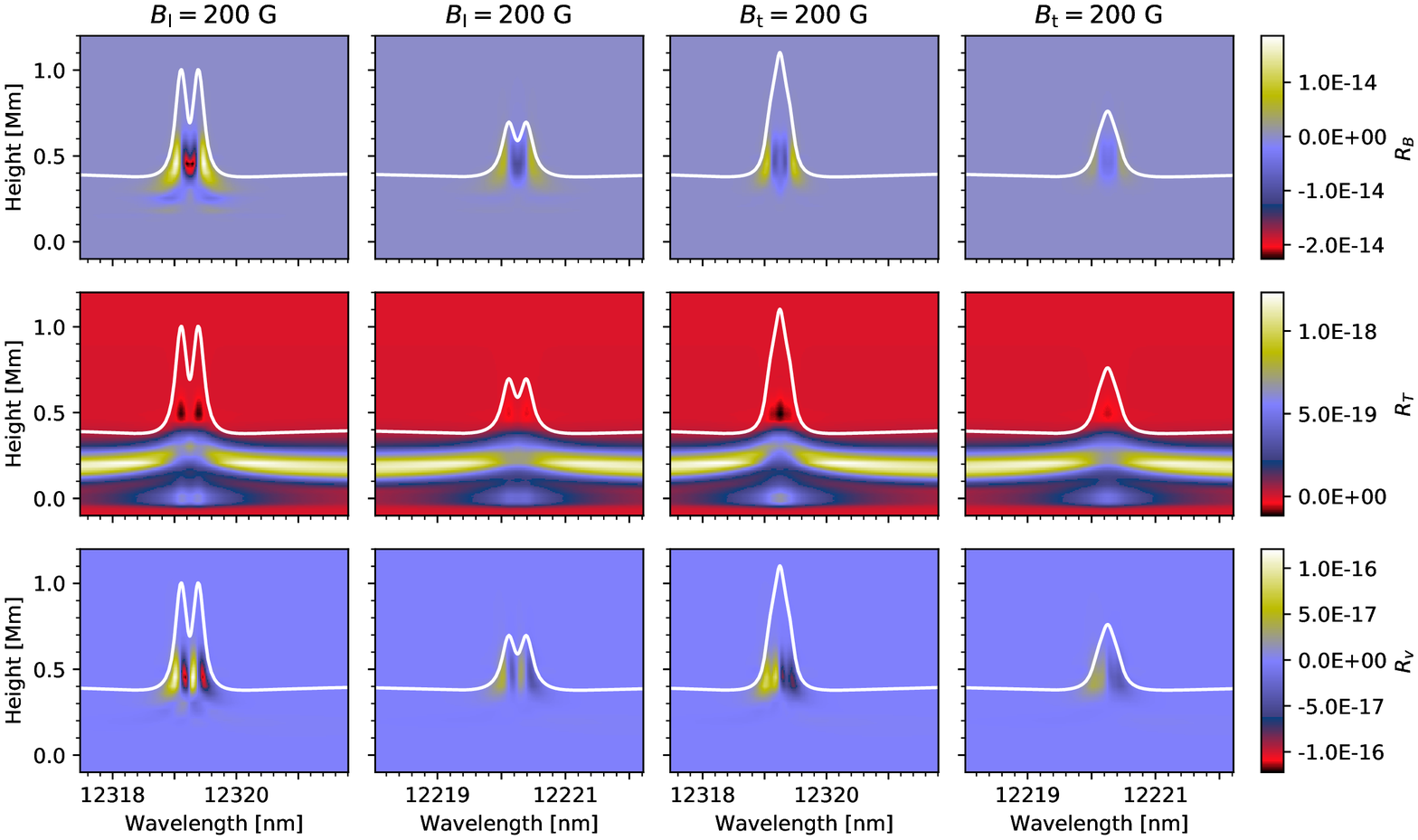} \caption{Response functions of Stokes I of the two Mg I 12 $\mu$m lines for the FALC model to magnetic field (up), temperature (middle), and velocity (bottom). Columns 1 and 3 and 2 and 4 correspond to the Mg I 12.32 $\mu$m line and the Mg I 12.22 $\mu$m line, respectively. The left two columns and the right two columns correspond to $B_l$ = 200 G and $B_t$ = 200 G, respectively. The white lines are the corresponding Stokes I profiles relative to the continuum intensity.}
\label{fig4}
\end{figure*}

\begin{figure*}[!htbp]
\centering
\includegraphics[width=18cm, angle=0]{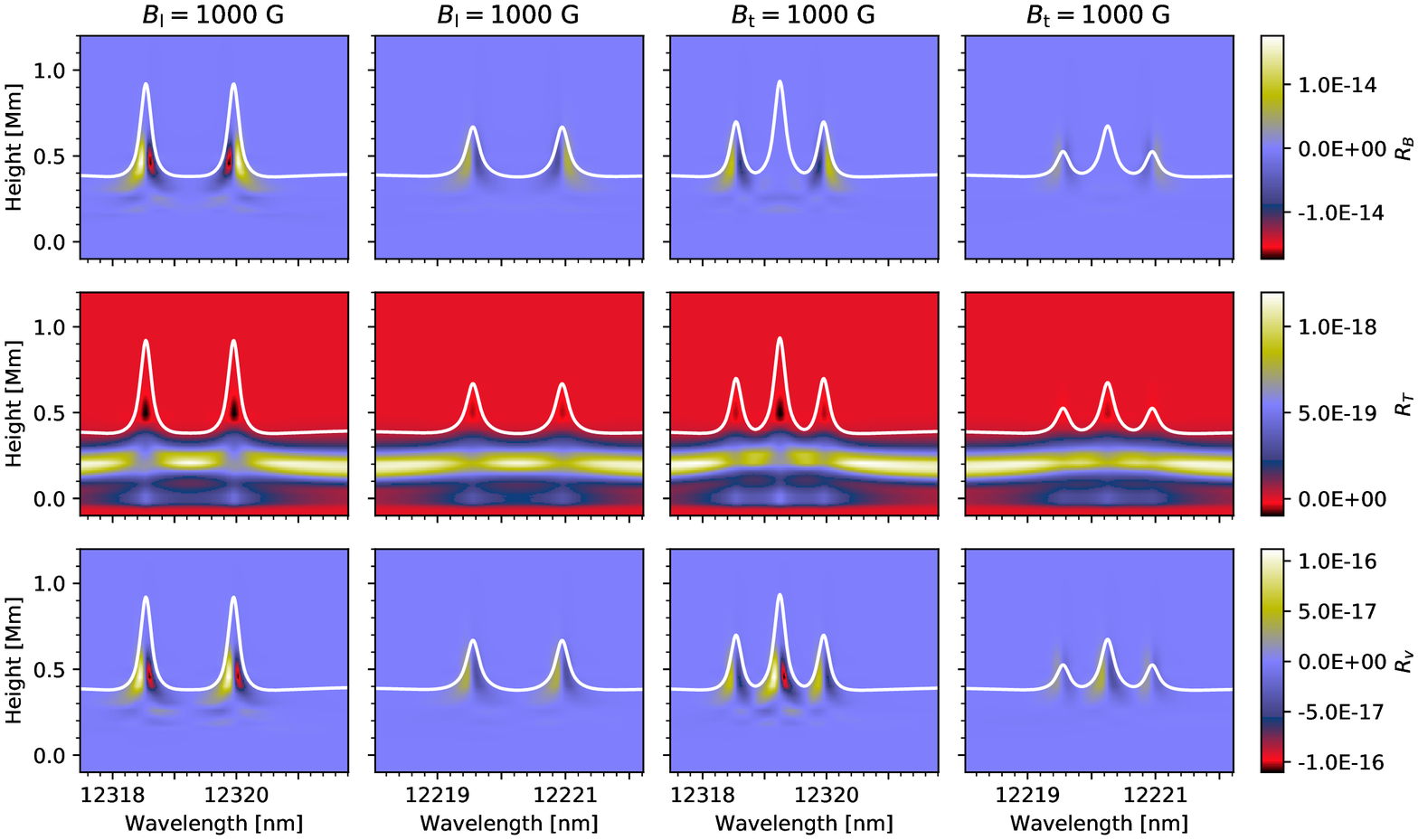} \caption{Same as Fig. 4, but the left two columns and the right two columns correspond to $B_l$ = 1000 G and $B_t$ = 1000 G, respectively.}
\label{fig5}
\end{figure*}

When physical parameters (e.g., magnetic field, temperature, and Doppler velocity) are diagnosed with a given spectral line, we need to know at which height they are mainly affected. The height information can be obtained based on the response function, which indicates how the Stokes profiles respond to small changes in the various physical parameters at different atmospheric heights \citep{1977Landi}. A response function with a high value means that the investigated physical parameter is more sensitive to that height. When different spectral lines are used to diagnose the same physical parameter (e.g., the magnetic field), the higher the response function value for the same disturbance, the more suitable this spectral line for diagnosing the physical parameter. Figures~\ref{fig4} and~\ref{fig5} show the response functions of Stokes I of the two Mg I 12 $\mu$m lines to the magnetic field, temperature, and velocity with $B_l$ and $B_t$. Here the magnetic field strengths of 200 G (Figure~\ref{fig4}) and 1000 G (Figure~\ref{fig5}) correspond to an incomplete and a complete split of the Stokes I profile, respectively.


The response function is a two-dimensional function of wavelength and height. The corresponding rows in Figures~\ref{fig4} and~\ref{fig5} show that the height corresponding to the maximum value of the response function of the Stokes I of the two Mg I 12 $\mu$m lines to perturbations in the magnetic field, temperature, and velocity at line center are 450 km, 490 km, and 450 km, respectively. However, the height corresponding to the maximum value of the response function of the Mg I 12.22 $\mu$m to perturbations in the velocity with $B_t$ = 200 G is only 400 km. This means that for the Mg I 12 $\mu$m lines, both the magnetic field and the velocity are mainly affected by the atmosphere around $\sim$450 km. Except for $B_t$ = 200 G, the velocity of 12.22 $\mu$m comes from a lower atmosphere than that of 12.32 $\mu$m. Of the three considered parameters, only the temperature affects the continuum. In the response function to the temperature, the continuum is sensitive to the atmosphere at a height of 203 km. The height corresponding to the temperature at line center is about 40 km higher than that corresponding to the magnetic field and the velocity. Thus, when Stokes I, Q, U, and V profiles are employed to derive different physical parameters, the temperature signature comes from a greater height than the magnetic field and the velocity.

With regard to the line splitting, the response function of Stokes I to perturbations in the magnetic field, the first rows of Figures~\ref{fig4} and~\ref{fig5} show that as the magnetic field strength increases, the Mg I 12 $\mu$m lines are from incompletely to fully split due to the Zeeman effect. As a result, the distance between the wavelengths corresponding to the maximum value of the response function also gradually increases as the Zeeman splitting increases. Therefore, the wavelength position corresponding to the greatest magnetic sensitivity separates with the line splitting.

When we compare columns 1 and 3 and columns 2 and 4 in Figures~\ref{fig4} and~\ref{fig5}, the response function of the 12.32 $\mu$m line to the magnetic field, temperature, and velocity is greater than that of the 12.22 $\mu$m line for the same disturbance. This means that the 12.32 $\mu$m line is more sensitive to these physical parameters than the 12.22 $\mu$m line. In addition, the first row of Figure~\ref{fig5} shows that the response to the magnetic field is obviously different for two $\sigma$ components in that they are oppositely antisymmetrical about the wavelength centers of the two components. In the third row of Figure~\ref{fig5}, the response to the velocity is the same in character for all three components as they all shift in the same direction by the same amount. Moreover, in the right two columns in Figure~\ref{fig5}, the response function of the Stokes I central $\pi$ component to the magnetic field is near zero because the $\pi$ component does not shift or broaden in the Zeeman effect. However, the response function of Stokes I central $\pi$ component to the temperature and the velocity appears stronger because the central $\pi$ component is taller, and thus has steeper wings, so that $\Delta$I is larger. The results of the response function show that the 12.32 $\mu$m line is more sensitive to the magnetic field, temperature, and velocity than the 12.22 $\mu$m line. Therefore, the Mg I 12.32 $\mu$m line is a better choice for measuring magnetic fields.
\subsection{Multiwavelength calibration curve between Stokes profiles and magnetic field}
When the Zeeman triple components are incompletely separated, the magnetic field strength cannot be directly inferred from Stokes I using the Zeeman-splitting formula. Polarization spectra are required instead. Based on the above analysis, in this section we adopt the wavelength-integrated method to infer $B_l$ and $B_t$ \citep{2008Lites}. Specifically, we integrate the Stokes profiles with the wavelength as follows to obtain the curve of the integrated area with the change of the magnetic field (i.e., the S-B curve). The different calculation methods (S1, S2, S3, and S4) are defined as
\begin{equation}\label{eq1}
  S_{1}=\frac{\vert \int_{\lambda_b}^{\lambda_0}V(\lambda)\mathrm{d}\lambda \vert + \vert \int_{\lambda_0}^{\lambda_r}V(\lambda)\mathrm{d}\lambda \vert}{I_c\int_{\lambda_b}^{\lambda_r}\mathrm{d}\lambda},
\end{equation}

\begin{equation}\label{eq2}
  S_{2}=\frac{\int_{\lambda_b}^{\lambda_r} \vert V(\lambda) \vert \mathrm{d}\lambda}{I_c\int_{\lambda_b}^{\lambda_r}\mathrm{d}\lambda},
\end{equation}

\begin{equation}\label{eq3}
  S_{3}=\int_{\lambda_b}^{\lambda_r} \left[\left(Q/I \right)^2 + \left(U/I \right)^2 \right]^{1/4} \mathrm{d}\lambda,
\end{equation}

\begin{equation}\label{eq4}
  S_{4}=\frac{\int_{\lambda_b}^{\lambda_r}  \left[ Q^2(\lambda) + U^2(\lambda) \right]^{1/2} \mathrm{d}\lambda}{I_c\int_{\lambda_b}^{\lambda_r}\mathrm{d}\lambda},
\end{equation}
where $\lambda_b$ and $\lambda_r$ represent the blue and red limits of the integration on the line ($\lambda_b=12315.233$ nm and $\lambda_r=12324.862$ nm for 12.32 $\mu$m, $\lambda_b=12215.557$ nm and $\lambda_r=12224.946$ nm for 12.22 $\mu$m), $\lambda_0$ represents the zero-crossing wavelength of the Stokes V profiles ($\lambda_0=12319.246$ nm for 12.32 $\mu$m, $\lambda_0=12220.251$ nm for 12.22 $\mu$m), and $I_c$ corresponds to the continuum intensity in each model. S1 and S2 describe the wavelength-integrated Stokes V (circular polarization) and are related to $B_l$. Their difference is that S1 divides the Stokes V profile into two parts and integrates separately, while S2 directly integrates the absolute value of the entire Stokes V profile. S3 is related to the wavelength-integrated Stokes Q and U and is the formula derived from the weak-field approximation \citep{1967Rayrole}. The integrand function of S3, multiplied by a simple linear coefficient, is generally used to obtain the transverse magnetic field in the calibration of filter-based magnetograph \citep{2004Su}. S4 represents the degree of linear polarization and also has a relation with $B_t$. It is worth mentioning that S1 and S4 are adopted for the magnetic field calibration of the quiet Sun by the Solar Optical Telescope / Spectro-Polarimeter onboard Hinode \citep{2008Lites}.

\begin{figure*}[!htbp]
\centering
\includegraphics[width=18cm, angle=0]{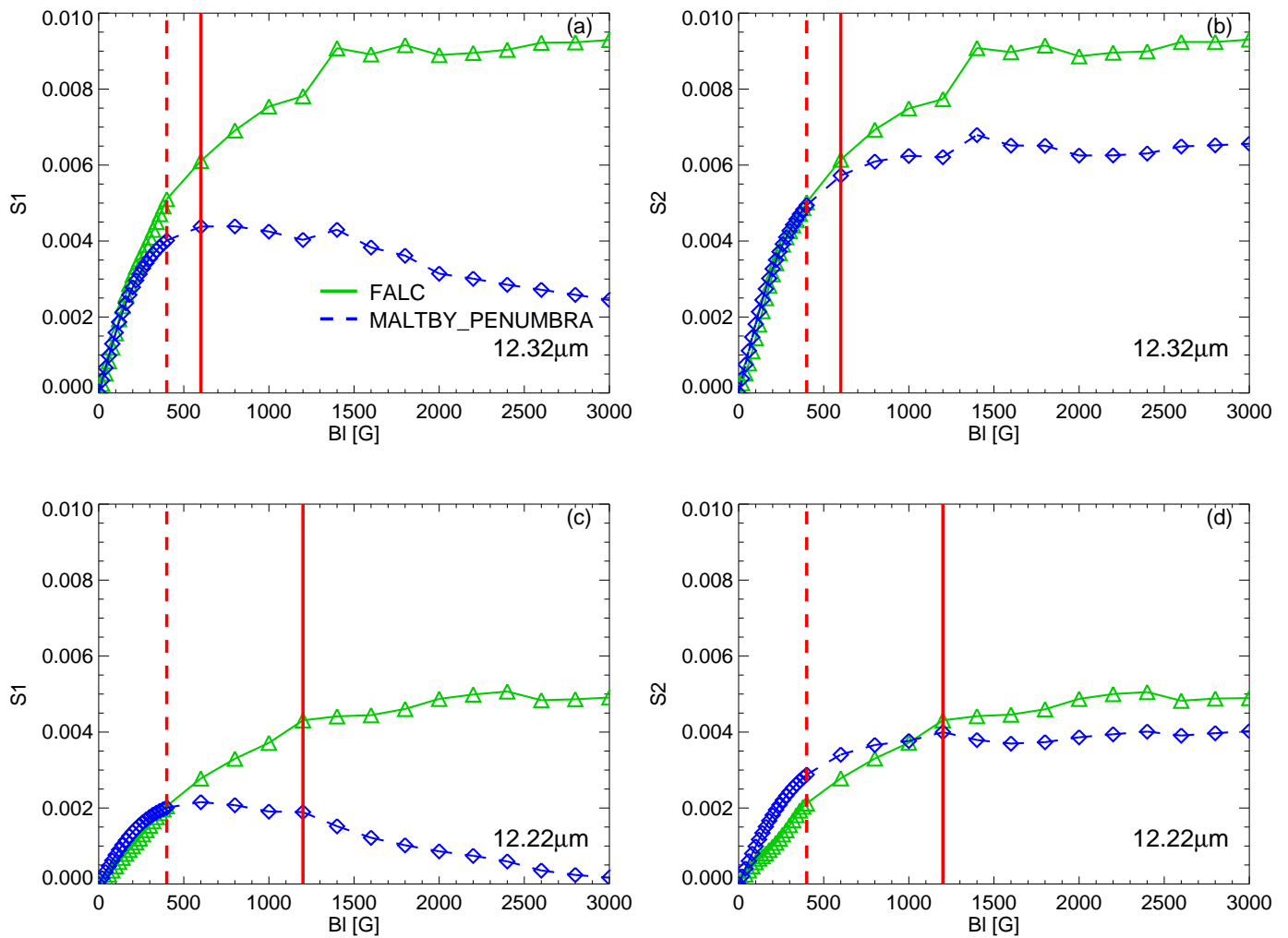} \caption{S-B curves of the Mg I 12.32 $\mu$m (top) and Mg I 12.22 $\mu$m (bottom) lines using different calculation methods in the FALC (solid green line, data points are denoted by triangles) and MALTBY (dashed blue line, data points are denoted by diamonds) models for $B_l$. S1 (left) and S2 (right) represent different wavelength-integrated methods (see Equations (1) and (2)). The vertical lines are used to divide the position that deviates from linearity in the FALC model (solid red) and MALTBY model (dashed red).}
\label{fig6}
\end{figure*}

\begin{figure*}[!htbp]
\centering
\includegraphics[width=18cm, angle=0]{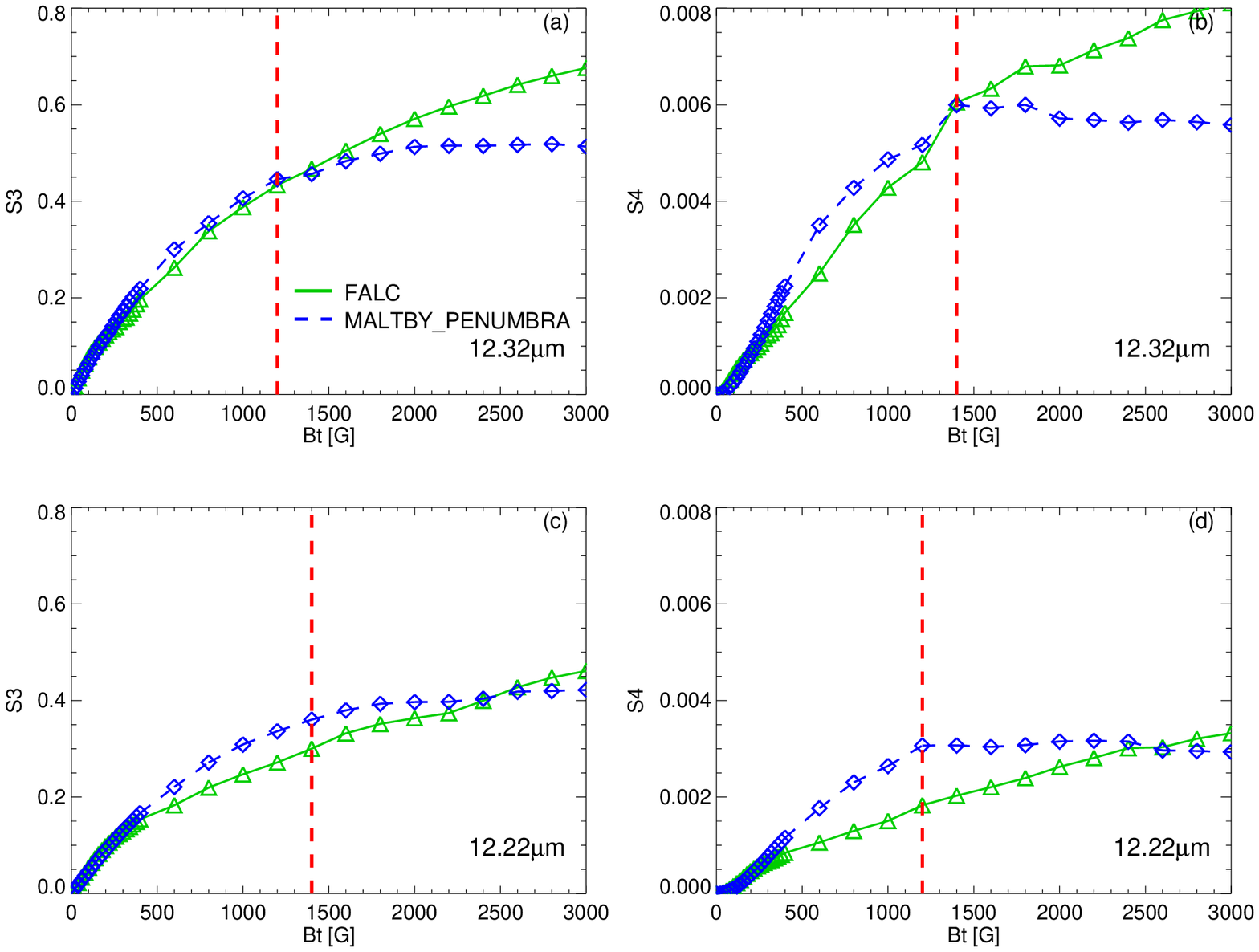} \caption{Same as Fig. 6, but for $B_t$. S3 (left) and S4 (right) represent different wavelength-integrated methods (see Equations (3) and (4)).}
\label{fig7}
\end{figure*}

Figures~\ref{fig6} and~\ref{fig7} display the calibration curves of S-$B_l$ and S-$B_t$, respectively, for the two Mg I 12 $\mu$m lines. When the magnetic field strength was weaker than 400 G, we chose 20 G as the interval in the calculation. The interval was set to 200 G for a magnetic field strength greater than 400 G. The calibration curves in the FALC model and the MALTBY model trend differently because the line width and model temperature are different. Furthermore, all the S-B curves are nonlinear distributions and just have a limited linear range for efficiently calibrating $B_l$ and $B_t$. We evaluated the four different methods from the overall trend of the curves, and the correlation coefficient (CC) was used as a reference or assistance to determine the approximate range for the linear calibration of $B_l$ and $B_t$. The vertical red lines in the figures roughly mark the upper limits of the linear range, the points within which have a CC value greater than 0.972.


Figure~\ref{fig6} shows that the two S-$B_l$ calibration curves (the S1 curve in the left panel and the S2 curve on the right) in the FALC model are almost the same for the two 12 $\mu$m lines. The S-$B_l$ curves in the MALTBY model show different behaviours after reaching saturation, that is, S1 gradually decreases, while S2 remains stable. One main reason is that S1 considers the effect of the wide absorption trough in the Stokes V profile. As the magnetic field strength increases, the wide absorption trough becomes deeper, so that the direct integration will cause the curve to decrease (cf., Appendix~\ref{fig9}). For the 12.32 $\mu$m (12.22 $\mu$m) line, both S1 and S2 have linear ranges of about 0--400 G (0--400 G) in the MALTBY model and about 0--600 G (0--1200 G) in the FALC model.

S-$B_t$ appears to have a broader linear range than S-$B_l$, as shown in Figure~\ref{fig7}. The calibration curves of both S3-$B_t$ and S4-$B_t$ in the FALC model are not saturated even at $B_t$ = 3000 G, showing good linearity within 0--3000 G. In the MALTBY model, S3 and S4 appear to have the linear ranges of about 0--1200 G (0--1400 G) and about 0--1400 G (0--1200 G), respectively, for the 12.32 $\mu$m (12.22 $\mu$m) line. Beyond the critical $B_t$ strengths indicated by the vertical lines, the dashed blue curves gradually become saturated. Neither model shows significant discrepancy between S3 and S4 for the two 12 $\mu$m lines.

These results show that the wavelength-integrated method is a very useful tool for calibrating $B_l$ and $B_t$ when the magnetic field is not strong enough to clearly see Zeeman splitting (weaker than 300G for $B_l$ and 500 G for $B_t,$ as shown in Figures~\ref{fig2} and~\ref{fig3}). The advantage of this method is that the signal-to-noise ratio can be improved by integrating the whole spectral line. Moreover, it is very fast by employing linear calibration compared with the Stokes inversion process. The disadvantage is that it can only diagnose a limited magnetic field range and has the saturation effect in strong magnetic fields. In the strong magnetic field case, we can use Zeeman-splitting formula and the ratio of the central $\pi$ to the $\sigma$ components to derive $B_l$ and $B_t$ (or the magnetic field strength and inclination angle) \citep{1990Deming}. The four different calculation methods considered are effective in calibrating $B_l$ and $B_t$. Therefore we can combine the S-B curves with the Zeeman triplet splitting to carry out a fast magnetic field calibration for both weak and strong magnetic field regions based on AIMS observations in the future. The Stokes inversion method considering the NLTE formation process of the Mg I 12 $\mu$m lines is still needed to derive the stratification of atmospheric parameters, that is, the height distribution of the temperature, magnetic fields, velocity, and so on.
\subsection{Single-wavelength calibration curve between Stokes profiles and magnetic fields}
A filter-based magnetograph is generally employed in the visible and near-infrared wavelength because it can obtain a large field-of-view magnetogram with high temporal resolution \citep{Schou2012,Deng2019,Tsuneta2008,Solanki2019,2010Cao}. Traditional filter-based magnetographs take routine Stokes observations just at one wavelength point of a selected magnetic sensitive line \citep{Ai1987,Hagyard1982}. In order to obtain the magnetic field, linear calibration under the weak-field approximation is generally adopted. In this section, we try to understand whether the single-wavelength magnetic field calibration method is suitable for the Mg I 12.32 $\mu$m line.

\begin{figure*}[!htbp]
\centering
\includegraphics[width=18cm, angle=0]{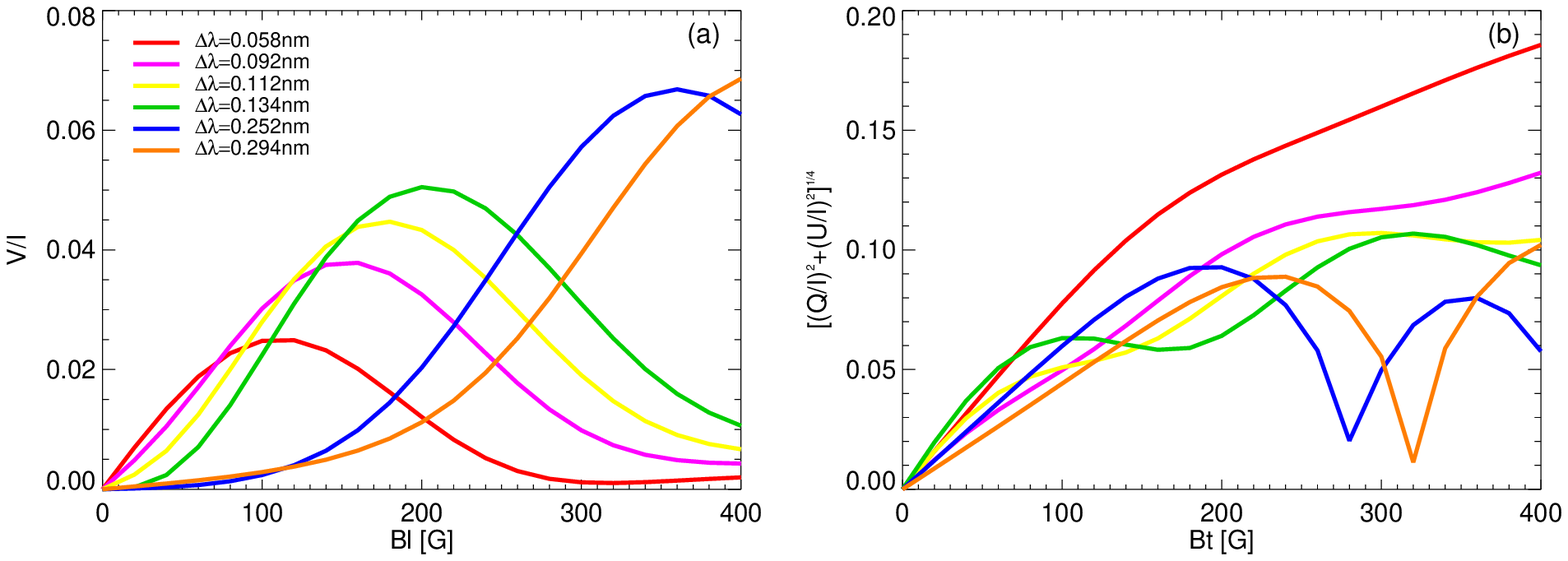} \caption{Single-wavelength calibration curves for different wavelengths. Panel (a) Calibration relation of the $B_l$ and Stokes V/I for the Mg I 12.32 $\mu$m line in the FALC model with a magnetic field strength 0--400 G. Panel (b): Same as panel (a), but the calibration relationship between $B_t$ and Stokes (Q/I, U/I). Different colours are for different wavelength positions toward the line center ($\lambda_0$ = 12319.246 nm). The selected wavelength positions are the same for $B_l$ and $B_t$.}
\label{fig8}
\end{figure*}

$B_{l}$ and $B_{t}$ were reconstructed by equations $B_{l}=C_{l}(V/I)$ and $B_{t}=C_{t}[(Q/I)^2+(U/I)^2]^{1/4}$, respectively, where $C_l$ and $C_t$ are the corresponding linear calibration coefficients. Figures~\ref{fig8}.a and~\ref{fig8}.b give the calibration curve of $B_{l}$ versus Stokes V/I and $B_{t}$ versus $[(Q/I)^2+(U/I)^2]^{1/4}$ for the FALC model with the magnetic field strength from 0 to 400 G, respectively. The calibration curves of V/I and $B_{l}$ strongly depend on the selected wavelength position. The farther the wavelength position is from the line center, the greater the saturation value of the magnetic field (Figure~\ref{fig8}a). These curves are approximately Gaussian distributed, which appears to reach saturation and then decreases. When the offset from the line center is 0.058 nm, 0.092 nm, 0.112 nm, 0.134 nm, 0.252 nm, and 0.294 nm, the corresponding value of magnetic saturation is about 100 G, 150 G, 180 G, 200 G, 360 G, and 400 G for the $B_{l}$. Because of the poor linearity of the calibration curves, the range of the magnetic field that can be diagnosed by linear calibration is very limited. Similar conclusions are found for $B_{t}$ (Figure~\ref{fig8}b). The calibration curves of $[(Q/I)^2+(U/I)^2]^{1/4}$ and $B_{t}$ have a quadratic distribution and a better linearity than $B_{l}$ if the offset $\Delta \lambda$ from the line center is less than 0.112 nm. However, the calibration curves are very complicated when the selected wavelength position $\Delta \lambda$ lies farther than 0.112 nm from the line center. So it is difficult to find a fixed wavelength position that can derive $B_{l}$ and $B_{l}$ with the linear calibration method even in the range of 0 to 400 G.

This result shows that the single-wavelength calibration method is not suitable for diagnosing magnetic fields in the Mg I 12.32 $\mu$m line because the effective linear range of the magnetic field is particularly small. As a comparison, the calibration curve using Fe I 532.4 nm line in the visible wavelength does not appear saturated until 1000 G when the linear calibration under the weak-field assumption is used \citep{1982Ai,2004Su}. This can be explained by the phenomenon that the Mg I 12.32 $\mu$m line usually shows triple Zeeman splitting with two or three hundred Gauss, while that for some visible spectral lines is about kilo-gauss. This means that the weak-field range of the Mg I 12.32 $\mu$m line is much smaller than that of visible lines with lower magnetic sensitivity.
\section{Conclusions and discussion}
We compared the differences in magnetic field diagnosis of the Mg I 12.32 $\mu$m and 12.22 $\mu$m lines based on the calculation results of the Stokes profile and response function, and calculated their calibration curve distribution using  wavelength-integrated methods. Moreover, we analyzed the feasibility of the single-wavelength calibration method using the Mg I 12.32 $\mu$m line. The main conclusions are listed below.

   \begin{enumerate}
      \item[(1)]
      The synthetic Stokes I profiles of the two Mg I 12 $\mu$m lines from the RH code without magnetic fields are consistent with previous observations and simulations. The radiation intensity of the 12.32 $\mu$m line is stronger than that of the 12.22 $\mu$m line, and the Saha-Boltzmann temperature sensitivity explains that there may be emission in sunspot umbrae.
      \item[(2)]
      Although their Land\'e g-factors are the same, it is easier to obtain data with high signal-to-noise ratio using the 12.32 $\mu$m line because its polarization signal is stronger than that of the 12.22 $\mu$m line for the same field. The Stokes Q and U profiles are more complex in quiet regions than in penumbrae because of the wider line width and the more pronounced absorption troughs in penumbrae.
      \item[(3)]
      According to the analysis of the response function, the 12.32 $\mu$m line is more suitable for diagnosing magnetic fields than the 12.22 $\mu$m line because its response function value to the magnetic field is higher. For the Mg I 12 $\mu$m lines, the derived temperature signal mainly comes from a height of about 490 km above the photosphere ($\tau_{500}$ = 1), while the magnetic field and velocity sensitivity correspond to a height of about 450 km.
      \item[(4)]
      The calibration curves of the magnetic field and wavelength-integrated Stokes profiles can be used for fast magnetic field calibration. In estimating $B_l$, the linear range is about 0--400 G for the MALTBY model for the two Mg I 12 $\mu$m lines, while it is about 0--600 G (0--1200 G) for the 12.32 $\mu$m (12.22 $\mu$m) line in the FALC model. In estimating $B_t$, the linear range is about 0--1200 G in the MALTBY model, while it does not show significant saturation up to $\sim 3$ kG for the FALC model for the two 12 $\mu$m lines.
      \item[(5)]
      Because of the high magnetic sensitivity in the Mg I 12.32 $\mu$m line, the effective linear range of the magnetic field calibration using a single-wavelength method is very limited. It is not suitable for a traditional filter-based magnetograph to measure the magnetic field at a single wavelength point.
   \end{enumerate}

All the analyses we presented are based on the simulation results of the forward model, and what we input is the intrinsic magnetic field strength, which does not change with height. We only considered the magnetized components, that is, assuming the filling factor is 1 (f=1). This is an ideal situation, which is difficult to achieve even for large solar telescopes. In the actual magnetic field observation, it is a reverse process from the Stokes parameter to the vector magnetic field. In this case, the filling factor must be considered. However, we cannot obtain information about the filling factor from the wavelength-integrated Stokes profiles. Therefore we give the calibration curve distribution with filling factors of 0.2, 0.4, 0.6, and 0.8 for different calculation methods (Equations (1)-(4)) in Appendix~\ref{fig10}. The wavelength-integrated method can only derive an approximate magnetic field strength. The Stokes inversion is still needed when an accurate value of the magnetic field is to be obtained. This needs to be kept in mind when our method is used for magnetic field calibration, especially for weak fields.\\

In addition, we only used one-dimensional solar atmospheric models to investigate the radiative transfer process of Mg I 12 $\mu$m lines. In recent years, two-dimensional and three-dimensional magnetohydrodynamic models such as those from the Bifrost and MURaM codes have been used, which can reproduce quiet-Sun and sunspot features quiet well \citep{Gudiksen2011,2018Nbrega,2018Moreno,Rempel2014,Cheung2008,Chen2017,Mart2017}. With these models and the RH code, we can obtain the polarization profiles of the Mg I 12 $\mu$m lines at different solar features to investigate their spatial distribution and temporal evolution, which is helpful for us to better understand future observations from AIMS. We will therefore try to carry out these works in future papers.

\begin{acknowledgements}
      We are very grateful to the referee for the valuable comments that helped improve the manuscript. X.L. would like to thank Sihui Zhong and Xianyu Wang for helpful discussions. This work was supported by the National Natural Science Foundation of China under grants 11427901, 11873062, 11803002, 11973056, 12003051, 12073040, 11773038, U1731241, 11703042, the Chinese Academy of Sciences under grants XDA15320102, XDA15052200, XDA15320302, and grant 1916321TS00103201.
\end{acknowledgements}


\begin{appendix}
\section{}
\begin{figure*}[!htbp]
\centering
\includegraphics[width=18cm, angle=0]{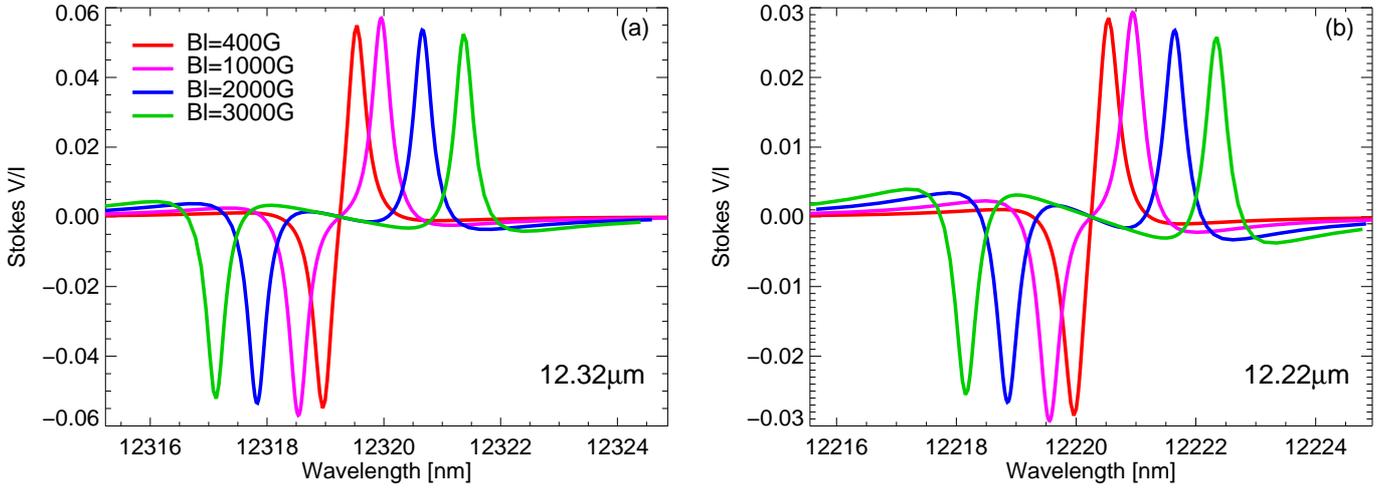}
\caption{The synthetic Stokes V profiles for different magnetic strengths. Panel (a) Calculated Stokes V profiles of the Mg I 12.32 $\mu$m line for sunspot penumbrae with $B_l$ = 400 G (red), $B_l$ = 1000 G (magenta), $B_l$ = 2000 G (blue), and $B_l$ = 3000 G (green). Panel (b): Same as panel (a), but for the Mg I 12.22 $\mu$m line.}
\label{fig9}
\end{figure*}

Figure~\ref{fig9} shows the Stokes V profiles for the MALTBY model with $B_l$ of 400, 1000, 2000, and 3000 G. For the two Mg I 12 $\mu$m lines, the wide absorption trough in the Stokes V profile becomes deeper as the magnetic field strength increases. When the magnetic field increases to 3000 G, the absorption trough becomes very obvious. This phenomenon is particularly clear for the 12.22 $\mu$m line.

\section{Multiwavelength calibration curve considering the filling factor}
\begin{figure*}[!htbp]
\centering
\includegraphics[width=18cm, angle=0]{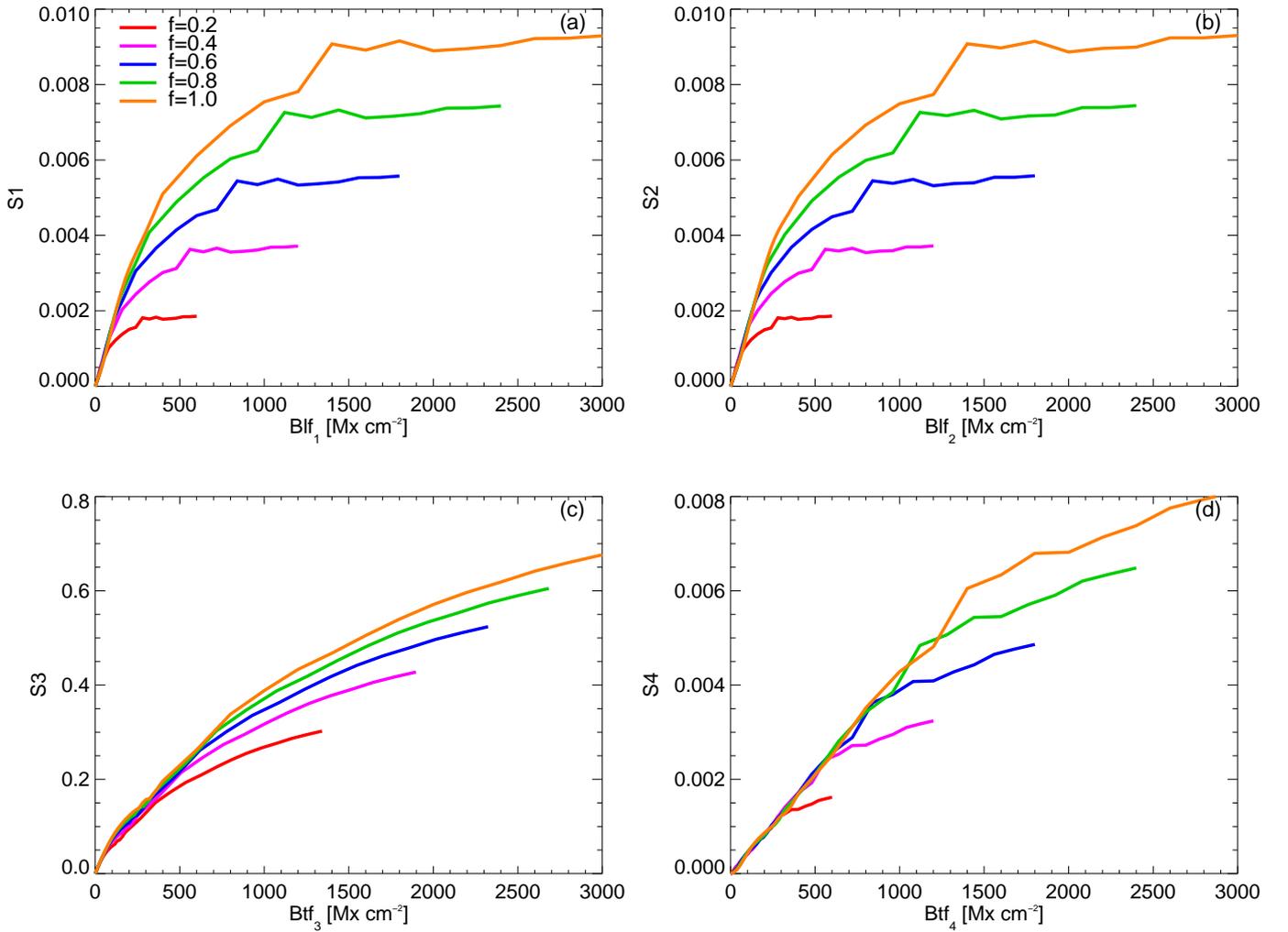}
\caption{In the FALC model, the multiwavelength calibration curves of the Mg I 12.32 $\mu$m line considering different filling factors for different calculation methods (see Equations (1)--(4)). Different colours show the results with different filling factors.}
\label{fig10}
\end{figure*}

Figure~\ref{fig10} shows the multiwavelength calibration curves with filling factors of 0.2, 0.4, 0.6, 0.8, and 1.0. After considering the filling factor, the abscissas $B_l$ and $B_t$ are replaced by the longitudinal ($B_{lf1}$, $B_{lf2}$) and transverse apparent flux density ($B_{tf3}$, $B_{tf4}$), respectively \citep{2008Lites}. Mx cm$^{-2}$, which is equivalent to gauss, is used as the unit of the apparent flux density. According to Equations (1)-(4), the relationship between apparent flux density and intrinsic magnetic field strength under the weak-field approximation is as follows \citep{2019Bellot}:

\begin{equation}\label{eq5}
\begin{split}
\begin{aligned}
  &B_{lf1} = B_{lf2} = fB_l,\\
  &B_{tf3} = \sqrt{f}B_t,\\
  &B_{tf4} = fB_t,
\end{aligned}
\end{split}
\end{equation}
where $B_{lf1}$ and $B_{lf2}$ and $B_{tf3}$ and $B_{tf4}$ are the corresponding longitudinal and transverse apparent flux density in Equations (1) and(2) and (3) and (4), respectively. Here $B_{tf3} = \sqrt{f}B_t$ is because Stokes Q and U are proportional to the square of the transverse apparent flux density. The corresponding relationship between the observed Stokes profiles and the real Stokes profiles can be expressed as \citep{1993Hewagama}

\begin{equation}\label{eq6}
\begin{split}
\begin{aligned}
  &I_{obs}=f*I_{mag}+(1-f)*I_{nonmag},\\
  &Q_{obs}=f*Q_{mag},\\
  &U_{obs}=f*U_{mag},\\
  &V_{obs}=f*V_{mag},
\end{aligned}
\end{split}
\end{equation}
where $I_{obs}$, $Q_{obs}$, $U_{obs}$, and $V_{obs}$ are the observed Stokes profiles, $I_{mag}$, $Q_{mag}$, $U_{mag}$,  and $V_{mag}$ are the magnetized components, and $I_{nonmag}$ is the part without magnetic fields. We combined equations (1)--(4), (B.1), and (B.2) to recalculate the calibration curve between the integrated area and the apparent flux density. The calibration curves for different filling factors are different. In the case of a weak field, the actual components of the magnetic field corresponding to the same apparent flux density will be quite different. In other words, we can only derive the apparent flux density, but not the actual components of the magnetic field. Therefore, the information of the filling factor is needed when the intrinsic magnetic field strength is to be obtained. Conversely, the calibration curves can provide reference information for the filling factor in the case of a strong field.
\end{appendix}
\end{document}